
%
%
%
\def\unlockat{\catcode`\@=11} 
\def\lockat{\catcode`\@=12}   
\unlockat
\def\m@ssage{\immediate\write16}  \m@ssage{}
\def\d@f@ult{}
%

%
%
\font\seventeenrm=cmr17

\font\twelverm=cmr12
\font\ninerm=cmr9
\font\sixrm=cmr6

\font\seventeenbf=cmbx12 at 17pt
\font\fourteenbf=cmbx12 at 14pt
\font\twelvebf=cmbx12
\font\ninebf=cmbx9
\font\sixbf=cmbx6

\font\seventeeni=cmmi12 at 17pt             \skewchar\seventeeni='177
\font\fourteeni=cmmi12 at 14pt              \skewchar\fourteeni='177
\font\twelvei=cmmi12                        \skewchar\twelvei='177
\font\ninei=cmmi9                           \skewchar\ninei='177
\font\sixi=cmmi6                            \skewchar\sixi='177

\font\seventeensy=cmsy10 scaled\magstep3    \skewchar\seventeensy='60
\font\fourteensy=cmsy10 scaled\magstep2     \skewchar\fourteensy='60
\font\twelvesy=cmsy10 at 12pt               \skewchar\twelvesy='60
\font\ninesy=cmsy9                          \skewchar\ninesy='60
\font\sixsy=cmsy6                           \skewchar\sixsy='60

\font\seventeenex=cmex10 scaled\magstep3
\font\fourteenex=cmex10 scaled\magstep2
\font\twelveex=cmex10 at 12pt

\font\ninex=cmex10 at 9pt
\font\sevenex=cmex10 at 9pt
\font\sixex=cmex10 at 6pt
\font\fivex=cmex10 at 5pt

\font\seventeensl=cmsl10 scaled\magstep3
\font\fourteensl=cmsl10 scaled\magstep2
\font\twelvesl=cmsl10 scaled\magstep1
\font\ninesl=cmsl10 at 9pt
\font\sevensl=cmsl10 at 7pt
\font\sixsl=cmsl10 at 6pt
\font\fivesl=cmsl10 at 5pt

\font\seventeenit=cmti12 scaled\magstep2
\font\fourteenit=cmti12 scaled\magstep1
\font\twelveit=cmti12

\font\seventeentt=cmtt12 scaled\magstep2
\font\fourteentt=cmtt12 scaled\magstep1
\font\twelvett=cmtt12

\font\seventeencp=cmcsc10 scaled\magstep3
\font\fourteencp=cmcsc10 scaled\magstep2
\font\twelvecp=cmcsc10 scaled\magstep1
\font\tencp=cmcsc10

\newfam\cpfam

\font\seventeenss=cmss17
\font\fourteenss=cmss12 at 14pt
\font\twelvess=cmss12
\font\tenss=cmss10
\font\niness=cmss9

\font\sevenss=cmss8 at 7pt
\font\sixss=cmss8 at 6pt
\font\fivess=cmss8 at 5pt
\newfam\ssfam
\newdimen\b@gheight             \b@gheight=12pt
\newcount\f@ntkey               \f@ntkey=0
\def\f@m{\afterassignment\samef@nt\f@ntkey=}
\def\samef@nt{\fam=\f@ntkey \the\textfont\f@ntkey\relax}
\def\rm{\f@m0 }
\def\mit{\f@m1 }         
\def\cal{\f@m2 }
\def\it{\f@m\itfam}
\def\sl{\f@m\slfam}
\def\bf{\f@m\bffam}
\def\tt{\f@m\ttfam}
\def\ssf{\f@m\ssfam}
\def\caps{\f@m\cpfam}
\def\seventeenpoint{\relax
    \textfont0=\seventeenrm          \scriptfont0=\twelverm
      \scriptscriptfont0=\ninerm
    \textfont1=\seventeeni           \scriptfont1=\twelvei
      \scriptscriptfont1=\ninei
    \textfont2=\seventeensy          \scriptfont2=\twelvesy
      \scriptscriptfont2=\ninesy
    \textfont3=\seventeenex          \scriptfont3=\twelveex
      \scriptscriptfont3=\ninex
    \textfont\itfam=\seventeenit    
    \textfont\slfam=\seventeensl    
      \scriptscriptfont\slfam=\ninesl
    \textfont\bffam=\seventeenbf     \scriptfont\bffam=\twelvebf
      \scriptscriptfont\bffam=\ninebf
    \textfont\ttfam=\seventeentt
    \textfont\cpfam=\seventeencp
    \textfont\ssfam=\seventeenss     \scriptfont\ssfam=\twelvess
      \scriptscriptfont\ssfam=\niness
    \samef@nt
    \b@gheight=17pt
    \setbox\strutbox=\hbox{\vrule height 0.85\b@gheight
                                depth 0.35\b@gheight width\z@ }}
\def\fourteenpoint{\relax
    \textfont0=\fourteencp          \scriptfont0=\tenrm
      \scriptscriptfont0=\sevenrm
    \textfont1=\fourteeni           \scriptfont1=\teni
      \scriptscriptfont1=\seveni
    \textfont2=\fourteensy          \scriptfont2=\tensy
      \scriptscriptfont2=\sevensy
    \textfont3=\fourteenex          \scriptfont3=\twelveex
      \scriptscriptfont3=\tenex
    \textfont\itfam=\fourteenit     \scriptfont\itfam=\tenit
    \textfont\slfam=\fourteensl     \scriptfont\slfam=\tensl
      \scriptscriptfont\slfam=\sevensl
    \textfont\bffam=\fourteenbf     \scriptfont\bffam=\tenbf
      \scriptscriptfont\bffam=\sevenbf
    \textfont\ttfam=\fourteentt
    \textfont\cpfam=\fourteencp
    \textfont\ssfam=\fourteenss     \scriptfont\ssfam=\tenss
      \scriptscriptfont\ssfam=\sevenss
    \samef@nt
    \b@gheight=14pt
    \setbox\strutbox=\hbox{\vrule height 0.85\b@gheight
                                depth 0.35\b@gheight width\z@ }}
\def\twelvepoint{\relax
    \textfont0=\twelverm          \scriptfont0=\ninerm
      \scriptscriptfont0=\sixrm
    \textfont1=\twelvei           \scriptfont1=\ninei
      \scriptscriptfont1=\sixi
    \textfont2=\twelvesy           \scriptfont2=\ninesy
      \scriptscriptfont2=\sixsy
    \textfont3=\twelveex          \scriptfont3=\ninex
      \scriptscriptfont3=\sixex
    \textfont\itfam=\twelveit    
    \textfont\slfam=\twelvesl    
      \scriptscriptfont\slfam=\sixsl
    \textfont\bffam=\twelvebf     \scriptfont\bffam=\ninebf
      \scriptscriptfont\bffam=\sixbf
    \textfont\ttfam=\twelvett
    \textfont\cpfam=\twelvecp
    \textfont\ssfam=\twelvess     \scriptfont\ssfam=\niness
      \scriptscriptfont\ssfam=\sixss
    \samef@nt
    \b@gheight=12pt
    \setbox\strutbox=\hbox{\vrule height 0.85\b@gheight
                                depth 0.35\b@gheight width\z@ }}
\def\tenpoint{\relax
    \textfont0=\tenrm          \scriptfont0=\sevenrm
      \scriptscriptfont0=\fiverm
    \textfont1=\teni           \scriptfont1=\seveni
      \scriptscriptfont1=\fivei
    \textfont2=\tensy          \scriptfont2=\sevensy
      \scriptscriptfont2=\fivesy
    \textfont3=\tenex          \scriptfont3=\sevenex
      \scriptscriptfont3=\fivex
    \textfont\itfam=\tenit     \scriptfont\itfam=\seveni
    \textfont\slfam=\tensl     \scriptfont\slfam=\sevensl
      \scriptscriptfont\slfam=\fivesl
    \textfont\bffam=\tenbf     \scriptfont\bffam=\sevenbf
      \scriptscriptfont\bffam=\fivebf
    \textfont\ttfam=\tentt
    \textfont\cpfam=\tencp
    \textfont\ssfam=\tenss     \scriptfont\ssfam=\sevenss
      \scriptscriptfont\ssfam=\fivess
    \samef@nt
    \b@gheight=10pt
    \setbox\strutbox=\hbox{\vrule height 0.85\b@gheight
                                depth 0.35\b@gheight width\z@ }}
%
%
\normalbaselineskip = 15pt plus 0.2pt minus 0.1pt 
\normallineskip = 1.5pt plus 0.1pt minus 0.1pt
\normallineskiplimit = 1.5pt
\newskip\normaldisplayskip
\normaldisplayskip = 15pt plus 5pt minus 10pt 
\newskip\normaldispshortskip
\normaldispshortskip = 6pt plus 5pt
\newskip\normalparskip
\normalparskip = 6pt plus 2pt minus 1pt
\newskip\skipregister
\skipregister = 5pt plus 2pt minus 1.5pt
\newif\ifsingl@    \newif\ifdoubl@
\newif\iftwelv@    \twelv@true
\def\singlespace{\singl@true\doubl@false\spaces@t}
\def\doublespace{\singl@false\doubl@true\spaces@t}
\def\normalspace{\singl@false\doubl@false\spaces@t}
\def\Tenpoint{\tenpoint\twelv@false\spaces@t}
\def\Twelvepoint{\twelvepoint\twelv@true\spaces@t}
\def\spaces@t{\relax
      \iftwelv@ \ifsingl@\subspaces@t3:4;\else\subspaces@t1:1;\fi
       \else \ifsingl@\subspaces@t3:5;\else\subspaces@t4:5;\fi \fi
      \ifdoubl@ \multiply\baselineskip by 5
         \divide\baselineskip by 4 \fi }
\def\subspaces@t#1:#2;{
      \baselineskip = \normalbaselineskip
      \multiply\baselineskip by #1 \divide\baselineskip by #2
      \lineskip = \normallineskip
      \multiply\lineskip by #1 \divide\lineskip by #2
      \lineskiplimit = \normallineskiplimit
      \multiply\lineskiplimit by #1 \divide\lineskiplimit by #2
      \parskip = \normalparskip
      \multiply\parskip by #1 \divide\parskip by #2
      \abovedisplayskip = \normaldisplayskip
      \multiply\abovedisplayskip by #1 \divide\abovedisplayskip by #2
      \belowdisplayskip = \abovedisplayskip
      \abovedisplayshortskip = \normaldispshortskip
      \multiply\abovedisplayshortskip by #1
        \divide\abovedisplayshortskip by #2
      \belowdisplayshortskip = \abovedisplayshortskip
      \advance\belowdisplayshortskip by \belowdisplayskip
      \divide\belowdisplayshortskip by 2
      \smallskipamount = \skipregister
      \multiply\smallskipamount by #1 \divide\smallskipamount by #2
      \medskipamount = \smallskipamount \multiply\medskipamount by 2
      \bigskipamount = \smallskipamount \multiply\bigskipamount by 4 }
\def\normalbaselines{ \baselineskip=\normalbaselineskip
   \lineskip=\normallineskip \lineskiplimit=\normallineskip
   \iftwelv@\else \multiply\baselineskip by 4 \divide\baselineskip by 5
     \multiply\lineskiplimit by 4 \divide\lineskiplimit by 5
     \multiply\lineskip by 4 \divide\lineskip by 5 \fi }
\interlinepenalty=50
\interfootnotelinepenalty=5000
\predisplaypenalty=9000
\postdisplaypenalty=500
\hfuzz=1pt
\vfuzz=0.2pt
%
%
%
\newif\ifr@duced
 \r@ducedfalse\Twelvepoint\m@ssage{Using 12pt 1 pg/sd.}
%
%
\Twelvepoint
\newdimen\fullhsize \newbox\leftpage
\newif\ifl@stp@geempty \l@stp@geemptyfalse
\ifr@duced
 \fullhsize=25cm \hsize=12cm \vsize=18cm
 \let\l@r=L
 \output={\almostshipout{\leftline{\vbox{
           \makeheadline\pagebody\makefootline}}}
         \advancepageno}
 \def\almostshipout#1{%
     \if L\l@r \count1=1
       \message{[\the\count0.\the\count1]}
       \global\setbox\leftpage=#1 \global\let\l@r=R
     \else \count1=2
       \shipout\vbox{\hbox to \fullhsize{\box\leftpage\hfil #1}}
       \global\let\l@r=L\fi}
\else
 \hsize=14cm \vsize=22cm   
\fi
\outer\def\bye{\par\vfill\supereject
    \ifr@duced\if R\l@r \l@stp@geemptytrue\null\vfill\eject\fi\fi
    \end}
\ifr@duced
  \hoffset=-7mm \voffset=-15mm
\else
  \hoffset=12mm \voffset=-6mm
\fi
%
%
\newif\ifletterstyle \newif\ifmemostyle \newif\ifpoemstyle
\newif\ifspanish \spanishfalse
\def\paperstyle{\letterstylefalse\normalspace}
\def\letterstyle{\letterstyletrue\singlespace\parindent=0pt
                 \advance\parskip by 2\parskip}
\def\storystyle{\letterstyletrue\singlespace
                 \advance\parskip by 1.5\parskip}

\newif\ifdr@ftmode \dr@ftmodefalse
\def\draftmode{\dr@ftmodetrue}
\def\comment#1{\ifdr@ftmode{\tt #1}\else\relax\fi}
\paperstyle  
%
%
\newif\iffrontpage \frontpagefalse
\newtoks\memofootline \newtoks\memoheadline \newtoks\paperfootline
\newtoks\letterfootline \newtoks\paperheadline \newtoks\letterheadline
\newtoks\date \newtoks\fecha \newtoks\d@te
\headline={\ifmemostyle\the\memoheadline%
           \else \ifletterstyle\the\letterheadline%
           \else \ifpoemstyle\hfil%
           \else \the\paperheadline\fi\fi\fi}
\paperheadline={\hfil\ifdr@ftmode{\sixtt Ceci n'est pas un preprint.}\fi\hfil}
\letterheadline={\ifnum\pageno=0 \hfil
  \else\ifspanish\rm p\'agina \ \folio\hfil\the\fecha
  \else\rm page \ \folio\hfil\the\date\fi\fi}
\memoheadline={\ssf Page \ \folio\hfil\the\date}
\footline={\ifl@stp@geempty\hfil%
\else\ifmemostyle\the\memofootline\else\ifpoemstyle\hfil\else%
\ifletterstyle\the\letterfootline\else\the\paperfootline\fi\fi\fi\fi}
\letterfootline={\hfil}
\memofootline={\hfil}
\paperfootline={\iffrontpage\hfil\else \hss\iftwelv@\twelverm\else\tenrm\fi
-- \folio\ --\hss \fi }
\def\monthname{\relax\ifcase\month 0/\or January\or February\or
   March\or April\or May\or June\or July\or August\or September\or
   October\or November\or December\else\number\month/\fi}
\def\today{\monthname\ \number\day, \number\year}
\date={\today}
\def\nombremes{\relax\ifcase\month 0/\or enero\or febrero\or
   marzo\or abril\or mayo\or junio\or julio\or agosto\or septiembre\or
   octubre\or noviembre\or diciembre\else\number\month/\fi}
\def\hoy{\number\day\ de\ \nombremes, \number\year}
\fecha={\hoy}
\def\Date{\ifspanish\d@te=\fecha\else\d@te=\date\fi
\line{\hfill\rm\the\d@te}\bigskip}

%
%
\skip\footins=\smallskipamount
\dimen\footins=24truecm
\newcount\fnotenumber
\def\clearfnotenumber{\fnotenumber=0} \clearfnotenumber
\def\fnote{\global\advance\fnotenumber by1 \generatefootsymbol
 \footnote{$^{\footsymbol}$}}
\def\fd@f#1 {\xdef\footsymbol{\mathchar"#1 }}
\def\generatefootsymbol{\iffrontpage\ifcase\fnotenumber
\or \fd@f 279 \or \fd@f 27A \or \fd@f 278 \or \fd@f 27B \else
\fd@f 13F \fi
\else\xdef\footsymbol{\the\fnotenumber}\fi}
\def\footnote#1{\edef\@sf{\spacefactor\the\spacefactor}#1\@sf
      \insert\footins\bgroup\singl@true\doubl@false
      \ifr@duced\ninepoint\else\tenpoint\fi
      \interlinepenalty=\interfootnotelinepenalty \let\par=\endgraf
        \leftskip=\z@skip \rightskip=\z@skip
        \splittopskip=10pt plus 1pt minus 1pt \floatingpenalty=20000
        \smallskip\item{#1}\bgroup\strut\aftergroup\@foot\let\next}
%
%
\newcount\secnumber \def\clearsecnumber{\secnumber=0}
\newcount\appnumber \def\clearappnumber{\appnumber=64}
\clearsecnumber \clearappnumber
\newif\ifs@c 
\newif\ifs@cd 
\s@cdtrue 
\def\unsectioned{\s@cdfalse\let\section=\subsection}
\newskip\sectionskip         \sectionskip=\medskipamount
\newskip\headskip            \headskip=8pt plus 3pt minus 3pt
\newdimen\sectionminspace
\ifr@duced
\sectionminspace=7pc
\else
\sectionminspace=10pc
\fi
\def\Titlestyle#1{\par\begingroup \interlinepenalty=9999
     \leftskip=0.02\hsize plus 0.23\hsize minus 0.02\hsize
     \rightskip=\leftskip \parfillskip=0pt
     \advance\baselineskip by 0.5\baselineskip
     \hyphenpenalty=9000 \exhyphenpenalty=9000
     \tolerance=9999 \pretolerance=9000
     \spaceskip=0.333em \xspaceskip=0.5em
     \iftwelv@\seventeenpoint\else\fourteenpoint\fi
   \noindent #1\par\endgroup }
\def\titlestyle#1{\par\begingroup \interlinepenalty=9999
     \leftskip=0.02\hsize plus 0.23\hsize minus 0.02\hsize
     \rightskip=\leftskip \parfillskip=0pt
     \hyphenpenalty=9000 \exhyphenpenalty=9000
     \tolerance=9999 \pretolerance=9000
     \spaceskip=0.333em \xspaceskip=0.5em
     \iftwelv@\fourteenpoint\else\twelvepoint\fi
   \noindent #1\par\endgroup }
\def\undertext#1{\vtop{\hbox{#1}\kern 1pt \hrule}}
\def\spacecheck#1{\dimen@=\pagegoal\advance\dimen@ by -\pagetotal
   \ifdim\dimen@<#1 \ifdim\dimen@>0pt \vfil\break \fi\fi}
\def\section#1{\cleareqnumber \s@ctrue \global\advance\secnumber by1
   \message{Section \the\secnumber}
   \par \ifnum\the\lastpenalty=30000\else
   \penalty-200\vskip\sectionskip \spacecheck\sectionminspace\fi
   \noindent {\caps\enspace\S\the\secnumber\quad #1}\par
   \nobreak\vskip\headskip \penalty 30000 }
\def\subsection#1{\par
   \ifnum\the\lastpenalty=30000\else \penalty-100\smallskip
   \spacecheck\sectionminspace\fi
   \noindent\undertext{#1}\enspace \vadjust{\penalty5000}}

\def\appendix#1{\cleareqnumber \s@cfalse \global\advance\appnumber by1
   \message{Appendix \char\the\appnumber}
   \par \ifnum\the\lastpenalty=30000\else
   \penalty-200\vskip\sectionskip \spacecheck\sectionminspace\fi
   \noindent {\caps\enspace Appendix \char\the\appnumber\quad #1}\par
   \nobreak\vskip\headskip \penalty 30000 }
\def\ack{\par\penalty-100\medskip \spacecheck\sectionminspace
   \line{\iftwelv@\fourteencp\else\twelvecp\fi\hfil ACKNOWLEDGEMENTS\hfil}%
\nobreak\vskip\headskip }
\def\refs{\begingroup \par\penalty-100\medskip \spacecheck\sectionminspace
   \line{\iftwelv@\fourteencp\else\twelvecp\fi\hfil REFERENCES\hfil}%
\nobreak\vskip\headskip \frenchspacing }
\def\endrefs{\par\endgroup}
%
%
\newskip\frontpageskip \frontpageskip=12pt plus .5fil minus 2pt
\newif\ifp@bblock \p@bblocktrue
\newif\ifm@nth \m@nthtrue
\newtoks\pubnum \pubnum={?}
\newtoks\pubtype \pubtype={\iftwelv@\twelvesl\else\tensl\fi\ (Draft)}
\newtoks\m@nthn@me
\newcount\Ye@r \advance\Ye@r by \year \advance\Ye@r by -1900
\def\Year#1{\Ye@r=#1}
\def\Month#1{\m@nthfalse \m@nthn@me={#1}}
\def\m@nthname{\ifm@nth\monthname\else\the\m@nthn@me\fi}
\def\titlepage{\global\frontpagetrue\paperstyle\hrule height\z@ \relax
               \ifp@bblock\pubblock\fi\relax }
\def\endtitlepage{\vfil\break\clearfnotenumber\frontpagefalse}
\def\nopubblock{\p@bblockfalse}
\def\pubblock{\line{\hfil\iftwelv@\twelverm\else\tenrm\fi%
QMW--PH--\number\Ye@r--\the\pubnum\the\pubtype}
              \line{\hfil\iftwelv@\twelverm\else\tenrm\fi%
{\tt hep-th/9410097}}
              \line{\hfil\iftwelv@\twelverm\else\tenrm\fi%
\m@nthname\ \number\year}}
\def\title#1{\vskip\frontpageskip\Titlestyle{\caps #1}\vskip3\headskip}
\def\author#1{\vskip.5\frontpageskip\titlestyle{\caps #1}\nobreak}
\def\andauthor{\vskip.5\frontpageskip\centerline{and}\author}

\def\address#1{\par\kern 5pt\titlestyle{
\it #1}}
\def\and{\par\kern 5pt \centerline{\sl and}}
\def\andaddress{\par\kern 5pt \centerline{\sl and} \address}

\def\abstract#1{\par\dimen@=\prevdepth \hrule height\z@ \prevdepth=\dimen@
   \vskip\frontpageskip\spacecheck\sectionminspace
   \centerline{\iftwelv@\fourteencp\else\twelvecp\fi ABSTRACT}\vskip\headskip
   {\noindent #1}}
\def\email#1{\fnote{\tentt e-mail: #1\hfill}}
\def\newaddress#1{\fnote{\tenrm #1\hfill}}
\let\thanks=\newaddress
%
%
\newcount\refnumber \def\clearrefnumber{\refnumber=0}
 \clearrefnumber
\newwrite\R@fs                              
\immediate\openout\R@fs=\jobname.references 
\def\closerefs{\immediate\closeout\R@fs} 
\def\refsout{\closerefs\refs
\catcode`\@=11                          
\input\jobname.references               
\catcode`\@=12			        
\endrefs}
\def\refitem#1{\item{{\bf #1}}}
\def\ifundefined#1{\expandafter\ifx\csname#1\endcsname\relax}
\def\[#1]{\ifundefined{#1R@FNO}%
\global\advance\refnumber by1%
\expandafter\xdef\csname#1R@FNO\endcsname{[\the\refnumber]}%
\immediate\write\R@fs{\noexpand\refitem{\csname#1R@FNO\endcsname}%
\noexpand\csname#1R@F\endcsname}\fi{\bf \csname#1R@FNO\endcsname}}
\def\refdef[#1]#2{\expandafter\gdef\csname#1R@F\endcsname{{#2}}}
%
%
\newcount\eqnumber \def\cleareqnumber{\eqnumber=0}
\newif\ifal@gn \al@gnfalse  
\def\veqnalign#1{\al@gntrue \vbox{\eqalignno{#1}} \al@gnfalse}
\def\eqnalign#1{\al@gntrue \eqalignno{#1} \al@gnfalse}
\def\(#1){\relax%
\ifundefined{#1@Q}
 \global\advance\eqnumber by1
 \ifs@cd
  \ifs@c
   \expandafter\xdef\csname#1@Q\endcsname{{%
\noexpand\rm(\the\secnumber .\the\eqnumber)}}
  \else
   \expandafter\xdef\csname#1@Q\endcsname{{%
\noexpand\rm(\char\the\appnumber .\the\eqnumber)}}
  \fi
 \else
  \expandafter\xdef\csname#1@Q\endcsname{{\noexpand\rm(\the\eqnumber)}}
 \fi
 \ifal@gn
    & \csname#1@Q\endcsname
 \else
    \eqno \csname#1@Q\endcsname
 \fi
 \ifdr@ftmode\rlap{\sixtt #1}\fi%
\else%
\csname#1@Q\endcsname\fi\global\let\@Q=\relax}
%
%
\newif\ifm@thstyle \m@thstylefalse \def\mathstyle{\m@thstyletrue}
\def\proclaim#1#2#3\par{\smallbreak\begingroup
\advance\baselineskip by -0.25\baselineskip%
\advance\belowdisplayskip by -0.35\belowdisplayskip%
\advance\abovedisplayskip by -0.35\abovedisplayskip%
\noindent\ifdr@ftmode\llap{\sixtt #3}\fi%
{\caps#1.\enspace}{#2}\par\endgroup%
\smallbreak}
\def\<#1>{\csname#1M@TH\endcsname}
\def\m@kem@th<#1>#2#3{%
\ifm@thstyle \global\advance\eqnumber by1%
 \ifs@cd%
  \ifs@c%
   \expandafter\xdef\csname#1M@TH\endcsname{{%
\noexpand #2\ \the\secnumber .\the\eqnumber}}%
  \else%
   \expandafter\xdef\csname#1M@TH\endcsname{{%
\noexpand #2\ \char\the\appnumber .\the\eqnumber}}%
  \fi%
 \else%
  \expandafter\xdef\csname#1M@TH\endcsname{{\noexpand #2\ \the\eqnumber}}%
 \fi%
 \proclaim{\csname#1M@TH\endcsname}{#3}{#1}\par%
\else%
 \proclaim{#2}{#3}{#1}\par%
\fi}%
%
%
\def\Thm<#1>#2{\m@kem@th<#1>{Theorem}{\sl#2}}
\def\Prop<#1>#2{\m@kem@th<#1>{Proposition}{\sl#2}}
\def\Def<#1>#2{\m@kem@th<#1>{Definition}{\rm#2}}
\def\Lem<#1>#2{\m@kem@th<#1>{Lemma}{\sl#2}}
\def\Cor<#1>#2{\m@kem@th<#1>{Corollary}{\sl#2}}
\def\Conj<#1>#2{\m@kem@th<#1>{Conjecture}{\sl#2}}
\def\Rmk<#1>#2{\m@kem@th<#1>{Remark}{\rm#2}}
\def\Exm<#1>#2{\m@kem@th<#1>{Example}{\rm#2}}
\def\Qry<#1>#2{\m@kem@th<#1>{Query}{\it#2}}
\def\Fact<#1>#2{\m@kem@th<#1>{Fact}{\sl#2}}
%
%

%
%
\def\leaderfill{\leaders\hbox to 1em{\hss.\hss}\hfill}
\def\boxit#1{\vcenter{\hrule\hbox{\vrule\kern8pt
      \vbox{\kern8pt#1\kern8pt}\kern8pt\vrule}\hrule}}

%
%
\def\ref#1{{\bf [#1]}}
\def\ie{{\it i.e.\/}}
\def\nl{\hfil\break}
%
%
\def\q@d{\vrule width 0.7em height 0.6em depth 0.2em}
\def\QED{\enspace\q@d}
\def\lapprox{\hbox{\lower3pt\hbox{$\buildrel<\over\sim$}}}
\def\gapprox{\hbox{\lower3pt\hbox{$\buildrel<\over\sim$}}}
\def\quotient#1#2{#1/\lower0pt\hbox{${#2}$}}
%
\def\to{\rightarrow}
%

%
%
%
\def\reals{{\bf R}} 
\def\integ{{\bf Z}} 
%
%
\def\Tr{\mathop{\rm Tr}}
\def\Res{\mathop{\rm Res}}
%
\def\underrightarrow#1{\vtop{\ialign{##\crcr
      $\hfil\displaystyle{#1}\hfil$\crcr
      \noalign{\kern-\p@\nointerlineskip}
      \rightarrowfill\crcr}}} 
\def\underleftarrow#1{\vtop{\ialign{##\crcr
      $\hfil\displaystyle{#1}\hfil$\crcr
      \noalign{\kern-\p@\nointerlineskip}
      \leftarrowfill\crcr}}}  

\def\comm#1#2{\left[#1\, ,\,#2\right]}
\def\pb#1#2{\left\{#1\, ,\,#2\right\}}
%
\def\pder#1#2{{{\partial #1}\over{\partial #2}}}
%
%
%
\def\fr#1/#2{\mathord{\hbox{${#1}\over{#2}$}}}
\def\Santiago{\address{%
   Departamento de F{\'\i}sica de Part{\'\i}culas Elementales\break
   Universidad de Santiago, Santiago de Compostela 15706, SPAIN}}
%
%
%
%
%

\def\NPB#1#2#3{{\sl Nucl. Phys.} {\bf B#1} (#2) #3}

\def\CMP#1#2#3{{\sl Comm. Math. Phys.} {\bf #1} (#2) #3}

\def\PLB#1#2#3{{\sl Phys. Lett.} {\bf #1B} (#2) #3}
\def\JMP#1#2#3{{\sl J. Math. Phys.} {\bf #1} (#2) #3}

\def\Invm#1#2#3{{\sl Invent. math.} {\bf #1} (#2) #3}
\def\LMP#1#2#3{{\sl Letters in Math. Phys.} {\bf #1} (#2) #3}
\def\IJMPA#1#2#3{{\sl Int. J. Mod. Phys.} {\bf A#1} (#2) #3}

\def\TMP#1#2#3{{\sl Theor. Mat. Phys.} {\bf #1} (#2) #3}

\def\JSM#1#2#3{{\sl J. Soviet Math.} {\bf #1} (#2) #3}
\def\MPLA#1#2#3{{\sl Mod. Phys. Lett.} {\bf A#1} (#2) #3}

\def\PJAS#1#2#3{{\sl Proc. Jpn. Acad. Sci.} {\bf #1} (#2) #3}
\def\JPSJ#1#2#3{{\sl J. Phys. Soc. Jpn.} {\bf #1} (#2) #3}

\def\AdiM#1#2#3{{\sl Annali di Matematica} {\bf #1} (#2) #3}

\def\SPD#1(#2)#3|{{\sl Sov. Phys. Docklady\/} {\bf #1} (#2) #3}
\def\CPAM#1(#2)#3|{{\sl Comm. Pure Appl. Math.\/} {\bf #1} (#2) #3}
\def\hepth/#1/{{\tt hep-th/#1}}
\lockat
%
%
%

\let\pb =\anticomm

\def\d{\partial}

\def\pdo{{\hbox{$\Psi$DO}}}

\def\comb[#1/#2]{\left[{#1\atop#2}\right]}

\def\rflecha#1{\setbox1=\hbox{\ninerm ~#1~}%
\buildrel\hbox{\ninerm #1}\over{\hbox to\wd1{\rightarrowfill}}}
\def\lflecha#1{\setbox1=\hbox{\ninerm ~#1~}%
\buildrel\hbox{\ninerm #1}\over{\hbox to\wd1{\leftarrowfill}}}

\def\Lalq{\Lambda_{\alpha,q}}

\def\ov{\over}
\def\fr#1/#2{\mathord{\hbox{${#1}\over{#2}$}}}

\def\med{\fr1/2}

\def\W{\mathord{\ssf W}}

\def\ope[#1][#2]{{{#2}\over{\ifnum#1=1 {z-w} \else {(z-w)^{#1}}\fi}}}


%
\refdef[Hall]{A. Capelli, C. Trugenberger and G. Zemba, \NPB{396}{1993}{465}
\nl S. Iso, D. Karabali and B. Sakita, \PLB{296}{1992}{143}. }
\refdef[TwoDGra]{M. Fukuma, H. Kawai, R. Nakayama, \CMP{143}{1991}{371};\nl
             H. Itoyama and Y. Matsuo, \PLB{262}{1991}{233}.}
\refdef[Fluid]{S. Nojiri, M. Kawamura, A Sugamoto, \MPLA{9}{1994}{1159},
and hep-th/9409164}
\refdef[Luki]{J. Gawrylczyk and J. Lukierski, University of Wroclaw preprint,
February 1993.}
\refdef[QCD]{D. Gross and W. Taylor, \NPB{356}{1991}{208}}
\refdef[INSTAN]{K. Takasaki, \CMP{94}{1984}{35};\nl
Q.-H. Park, \PLB{238}{1991}{287} }
\refdef[Hu]{H.L. Hu, \PLB{324}{1994}{293}}
\refdef[WReview]{P.~Bouwknegt and K.~Schoutens, {\sl ${\cal
W}$-Symmetry in Conformal Field Theory},  {\it Phys. Reps.} to
appear.}
\refdef[Univ]{J.~M.~Figueroa-O'Farrill and E.~Ramos,
\JMP{33}{1992}{833}.}
\refdef[Wn]{A.~B.~Zamolodchikov, \TMP{65}{1986}{1205};\nl
V.~A.~Fateev and S.~L.~Lykyanov, \IJMPA{3}{1988}{507}.}
\refdef[Dickey]{L.~A.~Dickey,  {\sl Soliton equations and Hamiltonian
systems},  Advanced Series in Mathematical Physics Vol.12,  World
Scientific Publ.~Co..}
\refdef[SoliHall]{H. Hiro-Oka and S. Saito: {\sl Quantum Hall effect from
soliton
equations}, hep-th/9312142}
\refdef[GD]{I.~M.~Gel'fand and L.~A.~Dickey, {\sl A family of
Hamiltonian structures connected with integrable nonlinear
differential equations}, Preprint 136, IPM AN SSSR, Moscow (1978).}
\refdef[Adler]{M.~Adler, \Invm{50}{1979}{403}.}
\refdef[KP]{E.~Date, M.~Jimbo, M.~Kashiwara, and T.~Miwa
\PJAS{57A}{1981}{387}; \JPSJ{50}{1981}{3866}.}
\refdef[WKP]{ J.~M.~Figueroa-O'Farrill, J.~Mas, and E.~Ramos,
\PLB{266}{1991}{298}; \nl
L.~A.~Dickey, {\sl Annals NY Acad.~Sci.} {\bf 491}(1987)
131}
\refdef[YuWu]{ F.~Yu and Y.-S.~Wu, \NPB{373}{1992}{713}.}
\refdef[WKPq]{J.~M.~Figueroa-O'Farrill, J.~Mas, and E.~Ramos
\CMP{158}{1993}{17}.}
\refdef[WinftyKP]{K.~ Yamagishi, \PLB{259}{1991}{436};\nl F.~Yu and
Y.-S.~Wu, \PLB{236}{1991}{220}}
\refdef[Radul]{A.~O.~Radul, in {\sl Applied methods of nonlinear
analysis and control}, pp. 149-157, Mironov, Moroz, and Tshernjatin,
eds.,  MGU 1987 (Russian).}
\refdef[BaKeKi]{I. Bakas, B. Keshin and E. Kiritsis:\CMP{151}{1993}{233}}
\refdef[Shubin]{M.~A.~Shubin, {\it Pseudodifferential Operators and
Spectral Theory}, Springer-Verlag 1987.}
\refdef[winfty]{I.~Bakas, \PLB{228}{1989}{406}; \CMP{134}{1990}{487}.}
\refdef[Winfty]{C.~N.~Pope, L.~J.~Romans, and X.~Shen,
\NPB{339}{1990}{191}.}
\refdef[Woneplusinfty]{C. N. Pope, L. J. Romans, and X. Shen,
\PLB{242}{1990}{401}.}
\refdef[Watanabe]{Y.~Watanabe, \LMP{7}{1983}{99};
\AdiM{136}{1984}{77}.}
\refdef[quantumwkp]{F.~Yu and Y.-S.~Wu, UU-HEP-92/11 ({\tt
hepth@xxx/9205117}).}
\refdef[BK]{I.~Bakas and E.~Kiritsis, Maryland/Berkeley/LBL preprint
UCB-PTH-P1/44,LBL-31213 and UMD-PP-92/37, Sept.1991.}
\refdef[DS]{V.~G.~Drinfel'd and V.~V.~Sokolov, \JSM{30}{1984}{1975}.}
\refdef[Lukier]{}
%

%
\overfullrule=0pt
\def\pubblock{ \line{\hfil\rm USC FT-17-94}
               \line{\hfil\tt hep-th/9410097}
               \line{\hfil\rm October 1994}}
\titlepage
\title{Centrally Extended $\W_{1+\infty}$ and the
KP Hierarchy}
\author{Fernando Mart\'\i nez-Mor\'as\email{fernando@gaes.usc.es}}
\andauthor{Javier Mas\email{jamas@gaes.usc.es}}
\Santiago
\abstract{It is well known that the centerless $\W_{1+\infty}$ algebra
provides a hamiltonian structure for the KP hierarchy. In this
letter we address the question whether the centerful version plays a similar
r\^ole in any related integrable system. We find that, surprisingly enough,
the centrally extended $\W_{1+\infty}$ algebra yields yet another Poisson
structure
for the same standard KP hierarchy.
This is proven by explicit construction of the infinitely many new
hamiltonians in closed form.}
\endtitlepage
\section{Introduction}
$\W_{1+\infty}$ is an ubiquitous mathematical structure. It appears in
totally
different
contexts. Most of them are intrinsically
2 dimensional: Quantum Hall effect \[Hall],  2-D quantum gravity \[TwoDGra],
2-D fluid dynamics \[Fluid], large N QCD \[QCD] etc.
 But also in different approaches to four dimensional quantum gravity
this algebra seems to play
 a relevant r\^ole
 \[INSTAN]\[Hu].

The content of this letter is mainly concerned with the KP hierarchy.
 Precisely the KP phase-space has proven to be the natural arena in the
construction
of $\W$ type algebras. It is an infinite dimensional
phase-space endowed with a (bi-)hamiltonian structure (see \[Dickey] for a
master's review), {\ie} there is
a pair of coordinated Poisson structures, where the so called ``first" is
linear
and the ``second" is non-linear. The former one was identified in \[WinftyKP]
as
the
centerless $\W_{1+\infty}$ algebra.

Inspired by this result there have been some attempts to see what
integrable system would arise from the centrally extended version of this
algebra
 \[Luki],
the natural conjecture being that the central extension should parameterize
some kind of (``quantum") integrable deformation of the KP hierarchy
and, thereafter,
of the KP equation. However, the
hard part of the job, namely: the construction of the infinite tower of
hamiltonians in involution with respect of these Poisson brackets was,
to our knowledge, not solved.
Hence the conjecture remained unproven.

In \[WKPq] it was shown that one need not restrict the KP phase space to
the ring of pseudodifferential operators
of the form $\d^q+ u_1\d^{q-1} + u_2 \d^{q-2}+...$ with $q\in \integ$. With
due
care many structures admit an analytic continuation to  complex
 values of $q$. This proved to be the case for the second Gelfand-Dickey
hamiltonian structure, and the Poisson-bracket algebra that one obtains
received the name of $\W_{KP}^{(q)}$.

Interestingly enough, this construction
showed how to recover the centrally extended $\W_{1+\infty}$ algebra
 as a particular contraction  $q\to 0$ of the algebra
$\W_{KP}^{(q)}$, thus
supporting  evidence that {\sl this algebra
 could provide again a hamiltonian structure for the
standard KP hierarchy}.
  In this letter we  prove that this is indeed the case by
 completing the analysis of \[WKPq]
 when $q \rightarrow 0$ ($q\in \reals_+$). After suitably
isolating some infinities that appear in the limiting procedure,
we manage to obtain all the
hamiltonians in closed form .

\section{The many phase-spaces of KP}

The KP hierarchy is defined as the infinite system of equations given in Lax
form
 by
$$
\pder{L}{t_n}=\comm{(L^n)_+}{L}~~~~ n=1,2,3,...., \(laxkp)
$$
where $L$ is the pseudodifferential operator
$ L=\d+u_2\d^{-1}+u_3\d^{-3}+...$, and $L_+$ and $L_-$ are the usual
 projections
onto differential and integral parts.
This system of equations is bi-hamiltonian, {\ie} it admits the form
of Hamilton's equations with respect to two different sets of Poisson
brackets
$$
\pder{L}{t_n}=
\pb{H_{n+1}}{L}_1=\pb{H_{n}}{L}_2, \(caca)
$$
The infinite set of hamiltonians can be expressed in closed form as follows
$$
 H_{n}= {1\ov n} \Tr L^n = {1\ov n} \int \Res L^{n} \(coco)
$$
where the residue $\Res$ picks the coefficient of $\d^{-1}$ in any $\pdo$.
The two set of Poisson brackets which have been labeled by 1 and 2 correspond
to the {\sl centerless} $\W_{1+\infty}$ and $W_{KP}$ algebras respectively
\[WinftyKP]\[WKP].

 The basic observation made in \[WKPq] is that one may implement the
KP hierarchy
on the space ${\cal S}_q$ of pseudodifferential operators ($\pdo$'s)
of the form
$$
\Lalq = \alpha \d^q + \sum_{j=1}^\infty ~ u_j \d^{q-j} \(qlaxop)
$$
where $\alpha$ and $q$ are complex numbers.
The use of non-integer powers of the derivative operator deserves
some explanation.
{}From the operational point of view, the only relevant piece of information is
contained in the composition law that generalizes the Leibnitz rule:
$$
\d^q f  = \sum_{j=0}^\infty \comb[q/j] f^{(j)} \d^{q-j} \(qleibnitz)
$$
involving the generalized binomial coefficients
$$
\comb[q/j] \equiv {q(q-1)\cdots(q-j+1)\over j!}~~~q\in \reals.\(genbin)
$$

Furthermore, we shall need to make sense of objects
like $\log \d$, which we will use later on. We choose to do so by
thinking about this operator as $\log \d = \lim_{q\to 0}\fr 1/q( \d^q -1) $,
and use
this limiting expression to extract the corresponding composition law from
\(qleibnitz)
$$
(\log\d) f(x) = f(x)\log\d - \sum_{j=1}^\infty {(-1)^j\ov j!}f^{(j)}(x)\d^{-j}
\(logd)
$$
Notice that for any {\pdo} $A\in {\cal S}_q$, the commutator
$[\log\d,A]$ is a \pdo\ in ${\cal S}_{q-1}$, {\ie} the commutator with
$\log\d$  lowers by one the order of $A$.

The Lax equations \(laxkp) can be implemented on ${\cal S}_q$
with due care. In order to do so we first need to define the
$q$'th root of $\Lalq$. $(\Lalq)^{1/q}$ can be obtained
from its generic expression with $q$ being an integer by formally allowing
the parameter $q$ to become an arbitrary complex number.
After a somewhat tedious calculation one obtains
$$\eqalign{
(\Lalq)^{1/q} = &  \alpha^{1/q} [\d + \fr 1/c u_1 +
               \fr 1/c (u_2 - \fr {q-1}/2 (u_1'+\fr 1/c u^2_1) )\d^{-1} \cr
    &  \fr 1/c ( u_3- \fr {q-1}/2 u_2' + \fr {q^2-1}/{12} u_1''
      -\fr {q-1}/c u_1u_2 +\fr {q(q-1)}/{2c} u_1u'_1
      +\fr {(q-1)(2q-1)}/{6c^2} u_1^3 ) \d^{-2}                          \cr
    &  \fr 1/c (u_4-\fr {q-1}/2 u_3'+\fr {q^2-1}/{12} u_2''
  - \fr {q^2-1}/{24} u_1''' \cr &
   - \fr {(q-1)(2q-1)(q+5)}/{24c} u'_1u'_1 - \fr {(q-1)(q+1)^2}/{12}
 u_1u''_1
  + \fr{q^2-1}/{2c} u_1'u_2 \cr & + \fr {q(q-1)}/{2c} u_1u'_2
   - \fr{q-1}/c u_1u_3 - \fr{q-1}/{2c} u_2^2
   - \fr{(q-1)(2q^2+q-1)}/{4c^2}  u_1'u_1^2 \cr & +\fr {(q-1)(2q-1)}/{2c^2}
      u_1^2 u_2
   - \fr{(q-1)(6q^2-5q+1)}/{24c^3} u_1^4 ) \d^{-3}  +...]
    }
 \(raizq)
$$
where $c$ stands for the product $\alpha q$.

Now the KP hierarchy is defined on the space ${\cal S}_q$ through the
following system
$$
\pder{\Lalq}{t_n}=\comm{((\Lalq)^{n/q})_+}{\Lalq}=
\comm{\Lalq}{((\Lalq)^{n/q})_-}~~,~~  n\in \integ  \(laxkpq)
$$
{}From the second form of these equations, it is evident that the field
$u_1$ does not evolve. Therefore, it is customary to choose as initial
conditions $u_1(x)=0$.

Also from \(raizq) it is obvious that the $n$-th equation is proportional to
$\alpha^{n/q}$. Therefore we may $renormalize$ all the times by defining
$\tilde t_n \equiv \alpha^{n/q} t_n$, so as to absorb this factor.
With these changes, the first few equations are given by
$$
\eqalign{
\pder{u_i}{\tilde t_1} =& u_i' ~~~~~ i=2,3....\cr
\pder{u_2}{\tilde t_2} =& 2 u'_3 - (q-2)u''_2 \cr
\pder{u_2}{\tilde t_3} =& 3 u'_4 + \fr {3(3-q)}/2 ) u''_3 +
\fr{(q-3)^2}/4 u'''_2  + \fr {3(3-q)}/c u_2u'_2  \cr
\pder{u_3}{\tilde t_2} =& 2 u'_4 + u''_3 -\fr {(q-1)(q-2)}/3 u'''_2
                          - \fr {2(q-2)}/c u_2u_2' \cr
\vdots & \cr
} \(eqmot)
$$
Notice that after rescaling the times, the factors $q$ and $\alpha$
always appear
in the combination $c=\alpha q$. This is an essential fact for the rest of
the discussion, and it can be seen to hold  for the whole hierarchy.
Moreover the actual value of $c$ is irrelevant and it can be made to dissapear
 by rescaling $u_i\to c u_i$. Nevertheless, we prefer to keep track
of this factor in what follows.

Following the usual steps, one may solve for the KP equation,
$$
{3\over 4}{\d^2 u_2\over \d \tilde t_2^2} = \pder{~}{x}
\left( \pder{u_2}{\tilde t_3} - {1\over 4} u_2''' - {3\over c} u_2 u_2'
\right)
{}~~~~~(x\equiv \tilde t_1) \(KPeq)
$$
It is a main result of \[WKPq] that for arbitrary values of $q\neq 0$,
the equations of motion in \(laxkpq) admit the form of Hamilton's equations,
$$
\pder{\Lalq}{\tilde t_n} = \pb{\tilde H_n}{\Lalq}_{2,q} \(equhamil)
$$
the Hamilton's functions being given by the general expression
$$
\tilde H_n = \fr c/n \alpha^{-n/q} \Tr (\Lalq)^{n/q}. \(hamexpres)
$$
As an example we write down explicitely the first three cases:
$$
\eqalign{
\tilde H_1^{(q)} = \int (&u_2 - \fr {q-1}/{2c} u_1^2) \cr
\tilde H_2^{(q)} = \int (&u_3 - \fr{q-2}/{c} u_1u_2 +\fr {(q-1)(q-2)}/{3c^2}
 u_1^3) \cr
\tilde H_3^{(q)} = \int (&u_4 - \fr{q-3}/{2c} u_1u'_2 + \fr{(q-1)(q-3)}/{8c}
 u_1u''_1
+ \fr{2q^2-9q+9}/{2c^2} u_1^2u_2 \cr
& -\fr{q-3}/c u_1u_3 - \fr {q-3}/{2c} u_2^2 - \fr{(q-1)(6q^2-27q+27)}/{24c^3}
u_1^4) \cr
} \(hamils)
$$
The brackets in \(equhamil) are a generalization of the (second)
Gelfand-Dickey brackets written in \(caca)
to the space  ${\cal S}_q$ \[WKPq].
 In terms of the basis functions $u_i, i=1,2,...$ these Poisson
brackets define a non-linear algebra named
$\W_{KP}^{(q)}$. Its first few brackets look as follows:
$$
\eqalign{
\pb{u_1(x)}{u_1(y)}_{2,q} = & c\d_x \cdot\delta(x-y) \cr
\pb{u_1(x)}{u_2(y)}_{2,q} = & (-c\fr{q-1}/2 \d^2 +(q-1)\d u_1)_x \cdot\delta
(x-y) \cr
\pb{u_1(x)}{u_3(y)}_{2,q} = & (c \fr{(q-2)(q-1)}/6 \d^3 - \fr{(q-1)(q-2)}/2
\d^2 u_1
                         + (q-2)\d u_2)_x \cdot\delta(x-y)   \cr
\pb{u_2(x)}{u_2(y)}_{2,q} = & (-c\fr{(q-1)(2q-1)}/6 ~\d^3 - u_2\d - \d u_2
                     +\fr{q(q-1)}/2 (\d^2 u_1-u_1\d^2) \cr &
                     + \fr{q(q-1)}/c u_1\d u_1~)_x \cdot\delta(x-y) \cr
\pb{u_2(x)}{u_3(y)}_{2,q} = & (c\fr{(q-1)(q-2)(3q-1)}/{24}~\d^4 +
 \fr{q(q-1)(q-2)}/6
(u_1\d^3-2\d^3 u_1)  \cr
 & +\fr{(q+1)(q-2)}/2 \d^2 u_2 - 2\d u_3 - u_3 \d
- \fr{q(q-1)(q-2)}/{2c} u_1\d^2 u_1 \cr & + \fr{q(q-2)}/c u_1\d u_2~)_x\cdot
\delta(x-y) \cr
\pb{u_2(x)}{u_4(y)}_{2,q} =& (c\fr{(q-1)(q-2)(q-3)(1-4q)}/{120} \d^5 -
u_4\d -3\d u_4  \cr
& + \fr{q(q-1)(q-2)(q-3)}/{24}(3\d^4u_1-u_1\d^4)  -\fr{(q-2)(q-3)(2q+1)}/6
\d^3u_2
\cr & +\fr{(q-3)(q+2)}/2 \d^2 u_3  + \fr{q(q-1)(q-2)(q-3)}/{3c} u_1\d^3 u_1
\cr & -\fr{q(q-2)(q-3)}/{2c} u_1\d^2u_2
+\fr{q(q-3)}/c u_1\d u_3~)_x \cdot \delta(x-y) \cr
\pb{u_3(x)}{u_3(y)}_{2,q} =& (-c\fr{(q-1)(q-2)(3q^2-6q+1)}/{60}\d^5
-(\d^2u_3-u_3\d^2)  \cr &  -2(u_4\d+\d u_4) + \fr{q(q-1)(q-2)}/6(\d^3 u_2 +
 u_2 \d^3)
\cr & - \fr{q(q-1)(q-2)(3q-5)}/8 (\d^4u_1
-u_1\d^4)  -\fr{q(q-1)(q-2)(2q-3)}/{6c} u_1\d^3 u_1 \cr & + \fr{q(q-2)}/c
 u_2\d u_2
 -\fr{q(q-1(q-2)}/{2c} (u_2\d^2 u_1 - u_1\d^2 u_2) \cr &
-\fr q/c(u_1\d u_3 + u_3\d u_1)~)_x\cdot \delta(x-y) \cr \vdots & \cr
} \(BraWKP)
$$
The reader may verify that with the information contained in \(hamils) and
in \(BraWKP), the expression \(equhamil) yields equations \(eqmot).

A word of caution here: the Poisson brackets \(BraWKP) that span the
 $\W_{KP}^{(q)}$
algebra do not stabilize the initial condition $u_1(x)=0$ except for the
Hamiltonian flows generated by \(hamils) that yield precisely the
KP-evolution
 equations
\(eqmot). Therefore in order to reproduce these equations correctly one has
to maintain the field $u_1$ throughout the calculation, and only at the very
end set it to zero.
One can however neglect terms with more that one $u_1$ in \(hamils), whose
contribution
to the equations of motion will be proportional to this field.
 Alternatively, one may set $u_1=0$ from the start, but
then the Poisson brackets need to be reduced consistently via the Dirac
procedure. The resulting non-linear algebra is named $\hat W_\infty^{(q)}$
\[WKPq],
(or, for $q=1$, $\hat W_\infty$ \[YuWu]).

\section{Full $\W_{1+\infty}$
 as a Hamiltonian Structure for KP at $q=0$ }

Let us have a closer look at equations \(eqmot). They become ill defined in
the
limit
$q\to 0$. However, due to the fact that in the denominator $q$ always
enters in
the combination $c=\alpha q$ we may define a more interesting ``scaling"
limit
where $q\to 0$ and $\alpha \to \infty$ such that $c=\alpha q$ is held
constant

No less important is the fact that this limit may be taken directly at the
level
of the Lax equations \(laxkpq); the Lax pair involving, on one side, the
following
operator
$$
 \Lambda_c = \lim_{q\to 0 \atop \alpha= c/q} \Lalq
= c\log \d + u_2 \d^{-2} +...     \(lambdace)
$$
 and, on the other, integer powers of its ``infinitesimal root"
(c.f.\(raizq)),
$$\eqalign{
(\Lambda_c)^{\infty} \equiv &  \lim_{q\to 0 \atop \alpha= c/q}~
\alpha^{-1/q} (\Lalq)^{1/q}  \cr
 = & \d +\fr 1/c u_2\d^{-1} +\fr 1/c(u_3-\med u'_2)\d^{-2} +
... \cr }
$$
Namely,  the Lax equation
$$
 \pder{~}{\tilde t_n}\Lambda_c = [((\Lambda_c)^{\infty})^n_+,\Lambda_c]
\(laxzero)
$$
automatically encodes all the limiting expressions obtained from
the flows in \(eqmot).
The term $c\log \d$ on the right hand side arises from $\lim_{q\to 0}\alpha
 \d^q$
and shows that  the Lax operator is peculiar when we induce the KP
 hierarchy
on
the space ${\cal S}_0$. Nevertheless, we should consider the Lax pair as an
auxiliary device and care only about the consistency of the system it defines.
In what concerns the KP equation \(KPeq), it survives this limit
intact.

A comment is in order. The relevant fact that the hierarchy defined by
the Lax system
\(laxzero) is no other than KP (yet in a peculiar basis),
can be proven using the formalism of Sato. In this language, the
KP flows rather live on the Volterra group of operators of the form
$\Phi = 1+a_1\d^{-1} + a_2\d^{-2} + ...$ defined as
$$
\pder{~}{\tilde t_n}
\Phi =-(\Phi \d^n\Phi^{-1})_-\Phi \(kpvolterra)
$$
Their commutativity follows as the result of a straightforward computation
 \[Dickey].
These flows can be induced on ${\cal S}_q$ by means of the {\sl dressing
transformation}
$ \Lalq = \Phi\alpha\d^q\Phi^{-1}$. On ${\cal S}_0$ we may as well represent
the flows if we dress instead $\log \d$:
$$
\Lambda_c = \Phi(c\log\d)\Phi^{-1} = c\log \d  + [\Phi,c\log\d]\Phi^{-1}
\(dress)
$$
{}From this expression, and using $\(kpvolterra)$,
 the Lax equation \(laxzero) is recovered.

One may wonder if the hamiltonian formulation of the KP flows expressed in
\(equhamil) is as robust as the Lax formulation in the limit $q\to 0$.
There are two separate pieces that we must check:
the Poisson brackets \(BraWKP) and the hamiltonians \(hamils).

Concerning the first piece, we again have to refer to \[WKPq] where this
 limit
 has
been shown to yield the famous centrally extended linear $\W_{1+\infty}$
algebra;
in short: $\lim_{q\to 0 \atop \alpha=c/q}\{~,~\}_{2,q} = \{~,~\}_{1+\infty}$
where
$$
\eqalign{
\{u_i(x),&u_j(y)\}_{1+\infty} =( c (-1)^{i+1} \fr {(i-1)! (j-1)!}/{(i+j-1)!}
 \d^{i+j-1} \cr
& - \sum_{l=1}^{j-1} \comb[j-1/l] \d^l u_{i+j-l-1} +
\sum_{l=1}^{i-1} \comb[i-1/l] u_{i+j-l-1}
(-\d)^l )_x \cdot \delta(x-y) \cr}  \(winftyc)
$$
The first few particular cases are easily recovered taking the limit directly
in \(BraWKP). It is important to note the role of $c$
{\sl that here parameterizes the central extension of the algebra}.

The question about the fate of the hamiltonian equations \(equhamil) in this
limit
can be recasted in a form that leads us back to the original motivation of
this work: {\sl using the Poisson brackets given by the centrally extended
algebra $\W_{1+\infty}$,
can we find related Hamiltonians for the KP hierarchy?.}

The answer looks trivially positive, as a glance at \(hamils) reveals no
patologies in the desired limit. More generally, using the freedom to rescale
$c=1$ ($i.e.~
\alpha = 1/q$),
 the limiting definitions  (cf. \(hamexpres) )
$$
\tilde H_n^{(0)} = \lim_{q\to 0}\fr {q^{n/q}}/{n} \Tr (\Lalq)^{n/q}.
\(hamilszero)
$$
yield well defined expressions for all $n$.
$$
\eqalign{
\tilde H_1^{(0)} = \int & u_2 \cr
\tilde H_2^{(0)} = \int &(u_3 + 2 u_1u_2 ) \cr
\tilde H_3^{(0)} = \int &(u_4 + \fr 3/2 u_1u'_2 + 3 u_1u_3 +\fr 3/2 u_2^2) \cr
 \vdots &\cr } \(hamilszeros)
$$
where we have discarded terms with higher powers of the field $u_1$ since,
eventually, they will not contribute to the equations of motion when we set
$u_1=0$.

As the flows generated by \(hamilszero) commute,
involution {\ie} $\{H_i,H_j\}_{1+\infty}=0$, follows automatically.

\section{Conclusions}

We would like to single out three concluding remarks:

1.- The centrally extended $\W_{1+\infty}$ algebra provides a hamiltonian
structure for the KP hierarchy.
The possibility of consistently defining the KP flows
at $q=0$ relies on the fact that all the singularities that appear can
be absorbed
in an infinite renormalization of the times $t_n \to \tilde t_n =
\alpha^{n/q}
t_n$ and the
hamiltonians $H_n \to \tilde H_n$.

2.- Contrary to some claims, we showed that the $c$ in
the central extension of the algebra does not parameterize
an integrable deformation of the KP hierarchy.

3.- A natural question to ask is: how about the centrally extended
 $\W_\infty$?
{\ie} for what hierarchy does this algebra provide a Poisson structure?
 The centerless case is, of course, not problematic since
it yields the first hamiltonian structure of KP at $q=1$ \[WinftyKP].
Concerning the centerful $\W_\infty$ one may wish to start
 from the centrally extended $\W_{1+\infty}$ and set $u_1=0$, but
then the Poisson
structure has to be consistently reduced via Dirac brackets. The reduced
algebra
{\sl is not} the linear $\W_\infty$ but instead a nonlinear algebra named
$\W_\infty^\#$ \[WKPq](else, first reducing
$u_1=0$ for $q\neq 0$ brings us from $W_{KP}^{(q)}$
to $\hat W_\infty^{(q)}$, and the subsequent contraction  $q\to 0$
yields back the same algebra).
It turns out that the centerful $\W_\infty$ algebra can be also produced
 out of $\W_{KP}^{(q)}$ as a different contraction, namely: $q\to 1$ and
$\alpha\to\infty$ with $c'=\alpha(q-1)$ kept finite; the central extension
being
 proportional to $c'$ in $W_\infty$.
 However in this limit $c=q\alpha \to \infty$, and a glance at
the equations of motion \(eqmot) reveals that the KP flows collapse since
all the nonlinear terms vanish. Of course, the KP equation \(KPeq)
linearizes
 as well. That is, within the present scheme,
the associated hierarchy is not KP but a linear truncation thereof.
Nevertheless
this does not rule that the centrally extended $\W_\infty$ algebra
could still be
a hamiltonian structure for KP, yet the construction of the hamiltonians
claims for a different approach and remains for the moment an open
question.

Summarizing,
we have shown how to express the KP hierarchy in hamiltonian form from the
centrally extended $\W_{1+\infty}$ algebra; in particular,
we have provided a closed expression for the hamiltonian functions.
 The method of analytic continuation in the parameter $q$ has
shown to be a powerful tool in our analysis. Notice,
for example, that standard
methods for obtaining the conserved charges from Lax equations,
like taking the traces of powers of the Lax operator, are not even
defined for operators of the form $c\log \d +.....$. Hence our
hamiltonians can only be defined via the limiting procedure expressed
in \(hamilszero).
The relevance of these hamiltonians in systems where the centrally extended
$\W_{1+\infty}$
algebra plays a dynamical r\^ole, such as the Quantum Hall Effect or D=2
non-critical
strings, is currently under investigation. 	It is also an appealing
 challenge
to search in them
for a physical counterpart of the deformation parameter $q$.

\ack
We are most thankful to E. Ramos for encouragement and discussions.
Also we would
like to thank
J.M. Figueroa O'Farrill for providing his {\sl Mathematica} package
{\sl Ring},
that
 handles with the calculus of pseudodifferential operators.
Finally we are grateful
to M. Fitzgibbon for a careful reading of the manuscript.

\refsout

\bye